# Teaching Network Storage Technology- Assessment Outcomes and Directions

## ABSTRACT
*The paper presents academic content, delivery and assessment mechanisms used, available resources including initial lessons from teaching Networked Storage Technology as a special topics course to students enrolled in two specific programs - IT and CS. The course is based on the EMC's vendor-neutral Storage Technology Fundamentals course. Furthermore, this manuscript provides a detailed review of how the course fits into our curriculum, particularly, how it helps achieving the 2008 ABET assessment requirements.*

## 1. INTRODUCTION
We are witnessing ascendance of storage technology into a position of a major part of IT infrastructure. After almost ten years of double digit increases in fielded storage [13, 23] pace seems to be increasing, see the upward revised estimate in march 2008 [2]. This is happening despite the relative stagnation in most other IT related investments in US after the burst of dot net economy excesses. Demand for new video (such as media imaging and surveillance) and audio data fuel the insatiable expansion of storage requirements beyond traditional corporate data bases and data warehouses. E-mails, e-commerce, etc. are just a few of the obvious business drivers, not to forget support for social networking and file sharing. Other related trends such as proliferation of analytical Intelligence systems and regulatory (Sarbanes-Oxley) requirements, combine with essential need to minimize downtime in order to assure business continuity. Modern disaster recovery strategies opened up reconsideration of tapes as media of choice for backups and archival storage, contributing to massively increased demand for online backups needed for quick restore and to explosive growth of demand for network disk storage in particular.

Technologically, gap between processors' speed and access time on individual disks is widening and network access to combined disk storage with its promise of high bandwidth to mask latency, lead to the rapid development of a set of distinct storage technologies commonly named Networked Storage Technology [6,7,11,13]. The Networked Storage Technology (NST) is becoming the most important resource in IT [13,17,24].

The Network Storage Industry specifically, EMC with its Storage Technology Certification and Academic Alliance Program [18] and NST related associations standardization efforts (SNIA with SMIS, DMTF with CIM, VMware), provided fair amount of shared common knowledge base for training of professionals and base of publicly available state of the art content to support educational programs. The expected demand for storage professionals by far outstrips [15] supply coming from academe.

The situation in academe is not particularly favorable, specifically in US, where it still seems to be shrinking the supply of talent (for example a compounded 14% annual decline in CS enrollments since 2000). The Computing Research Association states that CS majors continue to decline at a rapid rate [27]. This will limit potential supply for years to come even if trends in enrolment reverse this year it would take several years before students graduate. Given a substantial body of knowledge regarding sophisticated network storage technologies, simply adding weeks to existing coursework say in Databases (Backups), Networks (Storage Networking Protocols like FC, iSCSI etc.), Computer Architecture (I/O, Disks arrays- RAID etc.) and Servers OS (esp. virtualization aspect) while highly recommended, does not look convincing to industry, nor is it academically sufficient for the amount of new knowledge accumulated. An alternative in the form of a completely new course on Storage Technology seems more viable. The EMC, a leading player in SAN storage technology arena, provided its Storage Technology Fundamentals course via free-of-charge Academic Alliance to universities worldwide [22]. This paper reveals on how such a course can be used and fitted into technically oriented curricula in CS and IT. While our experience does not cover delivery to IS students after a trial run with IT and CS students we do not see many impediments to serve even larger body including students of IS, SE, and CE, that is any and all of the established program types under general umbrella of Computing or the IT in a broad sense, possibly even reaching some specialized, telecommunication oriented, degrees.

This paper presents our specific efforts on establishing such a course and properly situating it into our respective curricula in order to maximally contribute to the fulfillment of program expectations and preparations of our students to enter into NST profession in particular. The paper is organized as follows:
1) Introduction presenting the case for teaching a course on storage technology,
2) Course outcomes and their mappings to program expectations (outcomes and objectives in ABET terminology [1] and assessment instruments with student resources).
3) Course topical schedule, clearly indicating expanded coverage of academically required background topics,
4) Conclusion, addressing lessons learned from course delivery, alternative delivery formats and directions for future course improvement, and
5) References.

## 2. STORAGE TECHNOLOGY COURSE OUTCOMES AND ASSESSMENT PROCEDURES
Short course description:
Course covers fundamentals of modern storage infrastructure technology including NAS, SAN, DAS and CAS, in its application for Backups and Disaster Recovery.

CS NST course specific outcomes, upon completion of the course, students should be able to:
1. Describe storage technology solutions such as SAN, NAS, DAS and CAS,
2. Understand technologies and articulate available solutions to support an IT infrastructure including Business Continuity, Information Availability, Local and

Remote Replication, Backup and Recovery and Disaster Recovery needs of an organization,
3. Understand key tasks in successfully planning, deploying, managing, and monitoring a modern large data storage infrastructure,
4. Work in a team and quickly get up to speed with various proprietary technologies.
5. Identify contemporary storage virtualization technologies.

**Table 1. CS course objectives mapped to CS Program outcomes**

| CS Program Mission Objectives | Course Outcomes | How Measured |
|---|---|---|
| Provide a foundation in design, implementation, integration and testing of software systems integration | 1, 2, 3, 5 | Jeopardy-Quizzes, Cases |
| Promote the understanding of concepts that underlie computer science | 1, 2, 3, 5 | Labs, Cases |
| Provide experience with computer hardware systems | 1, 2, 3, 5 | Final Project Exam, Cases, Labs |
| Teach communication and interaction skills necessary for teamwork | 4 | Labs Reports, Presentation of Cases |
| Provide experience with practical and applied information technology | 5 | Labs, Cases, Final Exam |
| Prepare for jobs in the field of specialization | 1, 2, 3, 4, 5 | Labs, Cases, Final Exam |

IT NST course specific outcomes are:
1. Understand and have a working knowledge of storage technologies such as SAN, NAS, DAS and CAS,
2. Identify leading storage technologies that provide cost-effective IT solutions for medium to large scale businesses and data centers,
3. Understand important storage technologies' features such as availability, replication, scalability and performance,
4. Work in project teams to install, administer and upgrade popular storage solutions,
5. Identify and install current storage virtualization technologies,
6. Manage virtual servers and storage between remote locations.

**Table 2. IT course objectives mapped to IT Program outcomes**

| IT Program Objectives | Course Outcomes | How Measured |
|---|---|---|
| Demonstrate expertise in the core information technologies | 1, 2, 3, 5, 6 | Quizzes, Cases |
| Identify and define the requirements that must be satisfied to address the problems or opportunities faced by an organization or individual | 2, 3, 5 | Labs, Cases Final Exam |
| Identify and analyze user needs to design effective and usable IT-based solutions and integrate them into the user environment | 2, 4 | Final Exam, Quizzes |
| Demonstrate an understanding of best practices and standards and their application to the user environment | 1, 2, 3, 6 | Group Projects, Labs |
| Identify, evaluate and use current and emerging technologies and assess their applicability to address individual and organizational needs | 1, 2, 3, 5 | Labs, Cases, Final Exam |
| Create and implement effective project plans for IT-based systems | 2, 3 | Labs, Cases |
| Work effectively in project teams to develop and/or implement IT-based solutions | 1, 2, 4 | Group Projects, Cases, Labs |
| Communicate effectively and efficiently with clients, user and peers, both orally and in writing | 4 | Group Projects, Cases, Labs |
| Demonstrate independent critical thinking and problem solving skills | 2, 3, 4, 6 | Group Projects, Cases, Labs, Final Exam |

Note that course was cross-listed and offered to CS and IT groups simultaneously[1].

Besides traditional lectures, delivery mechanisms include the following:
   a) Invited Speakers (EMC, Oracle, Lefthand Networks, etc.)
   b) Case Studies (Microsoft Project Real, etc.)
   c) Hands on Labs (RAID, FreeNAS, NFS/SMB shares)

Industry relevant cases and technology product data sheets are presented in unbiased manner using a fairly large sample including references [in random order here] from IBM, Hitachi, HP, EMC, Isslion, ONStar, SUN, Brocade, Dell, Compliant, Pillar, Oracle, Microsoft, NetApp, CA, Coraid, etc. etc. The main concentration of references was drawn following up from reviews published by Storage magazine and extensive researching of the key terms on the internet, Storage Decisions 2007 conference attending vendors and background study of references listed at the end of this paper.

Assignments/Assessment Instruments include besides daily discussions and observations, and a supervised laboratory work, also take home exercises:

**1. Best six out of eight Labs-Cases with presentations (team): 60%**

    **Lab-1: RAID Lab report**
    **Lab-2: Present NAS Case**
    **Lab-3: Present SAN Case**
    **Lab-4: Present CAS Case**
    **Lab-5:** Virtualization hands on Lab report plus Presentation Cases \*\*\*
    **Lab-6: Present Backup Case**
    **Lab 7: Present Replication Case**
    **Lab 8: Present De-duplication Case**

---
[1] Official course listings are IT 5090 and CS 5090 Storage Network Technologies (Selected Topics)

**2. Three Jeopardy Quizzes [with class participation]: 10%**
**3. Final Project- Exam (individual SAN Challenge Project): 30%**

Student resources include over 400MB of class related supplementary materials, specifically for each assignment: examples of professional presentations, examples of real world implementations, and guidelines for mainly quantitative analysis. Furthermore all lecture extensions and supplemental readings are as well as the Storage Technology Fundamentals course materials (courtesy of EMC) are provided free of charge on a CD to students eliminating cost of textbook.

**Table 3. Consolidated (new ABET) Outcomes by content**

| Course Outcome | ABET General Computing Outcomes | ABET IT programs specific Outcomes | ABET CS programs specific Outcomes | Assessment |
|---|---|---|---|---|
| 1 | f, i | j, k, l, m | j, k | Labs, Cases |
| 2 | a, c | j, l | j | Labs, Cases |
| 3 | e | n | | Quizzes, Lab, Group Projects |
| 4 | d | k, l, n | j, k | Labs |
| 5 | i | j, k, l, m | j, k | Labs, Final |
| 6 | i | j, k, l, m | j, k | Labs |

## 2.1 ABET General Computing Program Outcomes:

a) An ability to apply knowledge of computing and mathematics appropriate to the discipline,
b) An ability to analyze a problem, and identify and define the computing requirements appropriate to its solution,
c) An ability to design, implement and evaluate a computer-based system, process, component, or program to meet desired needs,
d) An ability to function effectively on teams to accomplish a common goal,
e) An understanding of professional, ethical, legal, security, and social issues and responsibilities,
f) An ability to communicate effectively with a range of audiences,
g) An ability to analyze the local and global impact of computing on individuals, organizations and society, including ethical, legal, security and global policy issues,
h) Recognition of the need for, and an ability to engage in, continuing professional development,
i) An ability to use current techniques, skills, and tools necessary for computing practice.

## 2.2 ABET CS Specific program Outcomes:

j) An ability to apply mathematical foundations, algorithmic principles, and computer science theory in the modeling and design of computer-based systems in a way that demonstrates comprehension of the tradeoffs involved in design choices,
k) An ability to apply design and development principles in the construction of software systems of varying complexity.

## 2.3 ABET IT Specific Program Outcomes:

j) An ability to use and apply current technical concepts and practices in the core information technologies,
k) An ability to identify and analyze user needs and take them into account in the selection, creation, evaluation and administration of computer-based systems,
l) An ability to effectively integrate IT-based solutions into user environment,
m) An understanding of best practices and standards and their application,
n) An ability to assist in the creation of an effective project plan.

## 3. TOPICAL SCHEDULE[2]

Five weeks summer course: two professors (A-8 days and B-16 days)
Part A:
   Section 1 – Introduction and foundations
   *Day 1*
   Module 0.A- Course Orientation
       *Syllabus, Readings, Assignments and Grading*
   An invited speaker to discuss: Hiring trends to support Storage infrastructure
       *Module 1.1 – Meeting Data Storage Needs*
       *Module 1.2 – Data Center Infrastructure*
   *Days 2 and 3*
   *Module 1.3 – Storage Technology Trends*
   Module 0.B Theoretical and Hardware Fundamentals for Storage Networks
   - *Standardized CIM data model [required for graduate students]*
   - *Logical Network Diagramming Notation*
   - *Basic Elements of relevant Queuing Theory and Error Correcting codes*
   - *Performance analysis of underlining hardware technologies [I/O, cache etc.]*
   - *Review of course Assignments and available cases/readings*

   Section 2 - Storage Systems Architecture
   *Days 4 and 5*
       *Module 2.1 – The Host Environment*
       *Module 2.2 – Connectivity*
       *Module 2.3 – Physical Disks*
       *Module 2.4 – RAID Arrays*
   Lab-1 Selecting RAID
       *Module 2.5 – Disk Storage Systems*
   *Day 6*
   Module 0.C Review of Storage Network Protocols
   Module 0.D Security Considerations for Networked Storage

   Section 3 - Introduction to Networked Storage
   *Days 7 and 8*
       *Module 3.1 – Direct Attached Storage*
       *Module 3.2 - Network Attached Storage*
   Lab: NAS Case studies

   Part B:
   Section 3 - Introduction to Networked Storage

---
[2] **bolded** are main expansions to the original EMC course content

*Days 9, 10 and 11*
    *Module 3.3 – Storage Area Networks*
Lab SAN Cases

*Days 12 ,13 and 14*
    *Module 3.4  – IP SAN*
Lab : IP SAN Cases
    *Module 3.5 – Content Adressable Storage [CAS]*
Lab: CAS Cases

Section 4 – Information Availability
*Days  15 and 16*
    *Module 4.1 – Business Continuity Overview*
    *Module 4.2 – Back Up and Recovery*
Lab: Backup Cases
*Days  17 and 18*
    *Module 4.3 – Business Continuity Local*
    *Module 4.4 – Business Continuity Remote*
Module 0.E
- Applications of SAN RAID and other storage replication technologies in BC
- Storage replication with high availability of service
- Current physical limitations of remote storage systems.
Lab: Replication Cases

Section 5 – Managing and Monitoring
*Day 19*
    *Module 5.1 – Monitoring In the Data Center*
    *Module 5.2 – Managing In the Data Center*

Section 6 – Security and Virtualization
*Days 20 and 21*
    *Module 6.1 – Securing the Storage Infrastructure*
    *Module 6.2– Securing the Storage Infrastructure*
*Days 22, 23  and 24*
    **Invited speaker from Lefthand Networks: VSA, SAN/iQ and the future of iSCSI. Presented by John Easlick (Lefthand Networks)**
    ***Module 6.3- Trends in Virtualization and the Future of Storage***
**Lab: Building RAID 5 network storage.**
**Lab: Installing, configuring and managing FreeNAS storage.**
**Lab: Creating NFS/SMB network shares.**

*Day 25*
    Final Exam: Comprehensive Case Project

## 4. ADVANCED CONTENT
In order to further increase students' expertise, the content EMC content was expanded with the following topics:
- Performance and security analysis
- I/O System (Infiniband, etc.)
- Minimal elements of queuing theory
- Storage Networking Protocols (compared using OSI model)
- Common Information Model (storage domain)
- Logical Network Diagramming Notation SANDS
- Standard diagramming
- Storage Area protocols – iSCSI, iFCP, FCIP
- Business Continuity and Disaster Recovery using virtualization
- Storage and server virtualization

Server virtualization has rapidly advanced to the level where network storage systems must be available at 100% of time and provide reliability, manageability and scalability. Highly scalable systems save money in a long run but may require higher initial investments.  Although many companies claim high scalability, SAN administrators immediately know the limitations of deployed SANs as soon as storage space becomes inadequate. Popularity of VMware's Virtual Infrastructure 3 (VI3) enterprise solution progresses the advancement of virtualization of server to the next level – Distributed Resource Scheduling (DRS), High Availability (HA), Enhanced HA and Distributed Power Management (DMP). These are some of the essential technologies provided by VI3. One of the major components for VI3 is not just ESX machines and Virtual Infrastructure Management; it is the networked storage behind virtualization of servers. 100% availability of virtual servers is highly depended on the availability of SAN behind the ESX servers. LANs are usually viewed as the "front-end" networks whereas SANs are considered "back-end". Migration of virtual machines "live", without interruption of service is only possible if SAN is available at 100% of the time. The importance of network storage has been stressed many times; however, with technologies such as virtualization of servers, the rules of enterprise networking are constantly changing. Various Fibre Channel (FC) technologies for network storage exist today but the certain limitation such as high cost, scalability and proprietary hardware forced further advancement of Ethernet. IP-based systems provide great scalability and standards. Protocols such as FCIP, iFCP provide vast benefits to enterprise network storage systems but also have certain limitations. FC-based SANs are great when there is no need to extend SAN over a distance. As soon as the need for distance is involved, IP-based data provisioning takes place. Perhaps for those reasons, iSCSI evolved a protocol that does not involve utilization of any FC equipment (i.e. all Ethernet-based). iSCSI simply transports SCSI commands over TCP/IP. Ethernet-based systems are cost-effective and highly scalable systems but they are limited by the bandwidth of the Ethernet channel. Today, 10Gb/s standard is no longer the "golden" bandwidth – 40Gb/s and 100Gb/s transfer rates are soon to become new standards [9].

## 5. CONCLUSIONS
EMC Academic Alliance program is now active in seven countries - Brazil, China, India, Russia, Mexico, Ireland, and the US and reaching over 4000 students by 2008. This advanced Storage Technologies class had a tremendous success among the students that took the class. Many are ready to complete their storage certifications and a well-established communication venue with EMC allows graduates to fully meet the demands of the IT and CS job markets.
An established ABET course outcomes design and their mappings to program outcomes allowed instructors to stay focused and create active learning experience for students.

## 6. ACKNOLEDGEMENTS
Special thanks to Ed Van Sickle (EMC), Kimberly Yohannan (EMC), John Easlick (Lefthand Networks).